\documentclass[3p]{elsarticle}

\usepackage{amssymb}
\usepackage{graphicx}
          
\begin{document}

\begin{frontmatter}

\title{Model for tumour growth with treatment by continuous and pulsed 
chemotherapy}

\author{F. S. Borges$^1$, K. C. Iarosz$^1$, H. P. Ren$^2$, 
A. M. Batista$^3$, M. S. Baptista$^4$, R. L. Viana$^5$, S. R. Lopes$^5$, and 
C. Grebogi$^4$}
\address{$^1$Programa de P\'os-Gradua\c c\~ao em F\'isica, Universidade
Estadual de Ponta Grossa, 84030-900, Ponta Grossa, PR, Brazil} 
\address{$^2$Department of Information and Control Engineering, Xian 
University of Technology, 710048, Xian, PR, China}
\address{$^3$Departamento de Matem\'atica e Estat\'istica, Universidade
Estadual de Ponta Grossa, 84030-900, Ponta Grossa, PR, Brazil}
\address{$^4$Institute for Complex Systems and Mathematical Biology, 
SUPA, University of Aberdeen, AB24 3UE Aberdeen, United Kingdom}
\address{$^5$Departamento de F\'isica, Universidade Federal do Paran\'a,
81531-990, Curitiba, PR, Brazil}

\cortext[cor]{Corresponding author: $^3$antoniomarcosbatista@gmail.com}

\begin{abstract}
In this work we investigate a mathematical model describing tumour growth under 
a treatment by chemotherapy that incorporates time-delay related to the 
conversion from resting to hunting cells. We study the model using values for 
the parameters according to experimental results and vary some parameters 
relevant to the treatment of cancer. We find that our model exhibits a 
dynamical behaviour associated with the suppression of cancer cells, when 
either continuous or pulsed chemotherapy is applied according to clinical 
protocols, for a large range of relevant parameters. When the chemotherapy is 
successful, the predation coefficient of the chemotherapic agent acting on 
cancer cells varies with the infusion rate of chemotherapy according to an 
inverse relation. Finally, our model was able to reproduce the experimental 
results obtained by Michor and collaborators [Nature, 435, 1267 (2005)] about 
the exponential decline of cancer cells when patients are treated with the 
drug glivec.
\end{abstract}


\end{frontmatter}

\section{Introduction}

Cancer is the name given to a cluster of more than 100 diseases that presents a
common characteristic, the disorderly growth of cells that invade tissues and 
organs \cite{anderson,bru}. These cells may spread to other parts of the 
body rapidly forming tumours \cite{Baserga}.

An important mechanism of body defence against a disease caused by a virus, 
bacteria or tumour is the destruction of infected cells or tumours by actived
cytotoxic T-lymphocytes (CTL) cells also known as hunter lymphocytes. CTL are 
able to kill cells or to induce a programmed cell death (apoptosis). The 
biological activation process occurs efficiently when the CTL receive impulses 
generated by T-helper cells ($T_H$). The stimuli occur through the 
release of cytokines. This phenomenon is not instantaneous; besides 
the time elapsed to convert resting T-lymphocytes in CTL, there is  
also a natural delay of the cytological process \cite{wodarz,iarosz}. Banerjee 
and Sarkar studied the dynamical behaviour of tumour and immune cells using 
delay differential equations \cite{Sarkar2008}. They observed the existence of 
oscillations in tumour cells when a time delay was considered in the growth of 
T-cells.

A possible way to stop the growing of cancer cells is chemotherapy. That is, 
the treatment with a drug or combination of drugs through some protocol. There 
are many experimental and theoretical studies about the effects of the 
chemotherapy on the cells. Moreover, mathematical models have been considered 
to simulate the growth of cancer cells \cite{Liu}, as well as, tumour-immune 
interactions with chemotherapy \cite{Pillis}.

In this paper we investigate a mathematical model for the growth of tumours 
that not only take into consideration the time delay character of the 
lymphocytes dynamics, but also the effect of the chemotherapy. We extend the 
model of Sarkar and Banerjee \cite{Sarkar2008} by adding the chemotherapy, and 
by considering some clinically plausible protocols. Firstly, a continuous 
chemo\-therapy is analysed. Secondly, the traditional or pulsed chemotherapy 
protocol is analysed, in which the drug is administered periodically. 
According to experimental protocols, we have used both a constant amplitude 
\cite{Ahn} and an oscillatory amplitude \cite{Kuebler} for the continuous
infusion rate of chemotherapy \cite{Pinho2002}. 

One of our main results is to show that there are a large range of relevant 
parameters that lead to a successful chemotherapy. In a successful chemotherapy
is that the predation coefficient of the chemotherapic agent acting on the 
cancer cells and the infusion rate of the chemo\-therapy are inversely related.
For the continuous chemotherapy, we have ensured the stability of the 
non-cancer state (i.e., a successful chemotherapy) by calculating the Lyapunov 
exponents of the non-cancer solution. Finally, our model was able to reproduce 
the experimental results obtained by Michor and collaborators \cite{Michor} 
about the exponential decline of cancer cells when patients are treated with 
the drug glivec.

\section{The model}

We extend a mathematical model proposed by Banerjee and Sarkar 
\cite{Sarkar2005} including the chemotherapic agent.
The model is based on the predator-prey system. The T-lymphocyte is the
predator, while the tumour cell is the prey that is being attacked. The 
predators can be in a hunting or a resting state. The resting cells do not kill 
tumour cells,  but they can become hunters. The activation occurs not only due 
to cytokines released by macrophages that absorb tumour cells, but also by 
direct contact between resting and tumour cells. As a result, the resting cells
suffer a degradation while the hunting cells are actived. The activated cells 
do not return to the resting state. This way, the predator-prey model is a 
three dimensional deterministic system, consisting of tumour cells, hunting 
cells, and resting cells.
We added the chemotherapic agent in the equations as a predator on both 
cancerous and lymphocytes cells. 
The time delay of about 60 days considered in our model was observed by 
Balduzzi and collaborators \cite{Balduzzi,Villasana2003}, when they were 
realising experiments about lymphoblastic leukaemia. It incorporates many 
different phenomena in the system. It is one order of magnitude larger than the
one observed in Ref. \cite{Becker}. In our model, the time delay represents the 
total time interval for cancer cells to be identified by T-cell receptors and 
transfer this information to the killer cells, and the time related to the 
process of cytolytic information in the resting cells 
\cite{Becker,Matta}. The model is then given by 
\begin{eqnarray} 
\frac{dC(t)}{dt} &=& q_1 C(t) \left(1 - \frac{C(t)}{K_1}\right) - \alpha_{1} 
C(t) H(t) \nonumber\\
& & - \frac{p_1 C(t)}{a_1 + C(t)} Z(t), \nonumber\\
\frac{dH(t)}{dt} &=& {\beta}_1 H(t) R(t - \tau) - d_1 H(t) - \alpha_{2} C(t) 
H(t) \nonumber\\ 
& & - \frac{p_2 H(t)}{a_2 + H(t)} Z(t), \nonumber\\
\frac{dR(t)}{dt} &=& q_2 R(t) \left(1 - \frac{R(t)}{K_2}\right) - {\beta}_1 
H(t) R(t - \tau) \nonumber\\
& & - \frac{p_3 R(t)}{a_3 + R(t)} Z(t), \nonumber\\
\frac{dZ(t)}{dt} &=& \Delta - \left( \xi + \frac{g_1 C(t)}{a_1 + C(t)} + 
\frac{g_2 H(t)}{a_2 + H(t)} \right. \nonumber\\ 
& & \left.+ \frac{g_3 R(t)}{a_3 + R(t)}\right) Z(t), 
\label{modelo}
\end{eqnarray}
where $C$, $H$ and $R$ are the number of cancerous, hunting and resting cells,
respectively, $t$ is the time and $Z$ is the concentration of the chemotherapic
agent. The cancerous and resting cells have a logistic growth. The term 
$-d_1H(t)$ represents the natural death of the hunting cells. The terms 
$-\alpha_1C(t)H(t)$ and $\alpha_2C(t)H(t)$ are the losses due to encounters
between the cancerous and hunting cells. The term 
$\beta_1 H(t) R(t - \tau)$ is associated with the conversion of resting to 
hunting state, where $\tau$ is the delay in the conversion. The terms with
$Z$ correspond to interaction of the chemotherapic agent with the cells.

Table \ref{tab1} shows the parameters obtained from the literature,
according to experimental evidence, and Table \ref{tab2} shows the definition 
of some of the parameters. Table \ref{tab3} presents the values that we 
consider in our simulations for the sake of numerical integration.

\begin{table}[hbt]
\centering
\caption{Parameters according to experimental evidence.}
\begin{tabular}{ c | p{2.9cm} | c | c }
\textbf{\small Parameter}& \textbf{\small Definition}& \textbf{\small Value}& 
\textbf{\small Ref.}\\ \hline
{\footnotesize $q_1$}&{\footnotesize growth rate of malignant tumour cells}&
{\footnotesize 0.18 day$^{-1}$}&{\footnotesize \cite{Siu1986}}\\
{\footnotesize $K_1$}&{\footnotesize carrying capacity of tumour cells}&
{\footnotesize 5 x ${10}^6$ cells}&{\footnotesize \cite{Siu1986} }\\
{\footnotesize ${\alpha}_1$}&{\footnotesize decay rate of tumour}&{\footnotesize 1.101 x ${10}^{-7}$}&{\footnotesize \cite{Kusnetsov1994}}\\
{}&{\footnotesize  cells by hunting cells}&{\footnotesize  cells$^{-1}$ day$^{-1}$}&{}\\
{\footnotesize ${\alpha}_2$}&{\footnotesize decay rate of hunting }&{\footnotesize 3.422 x ${10}^{-10}$}&{\footnotesize \cite{Kusnetsov1994}}\\ 
{}&{\footnotesize cells by tumour cells }&{\footnotesize  cells$^{-1}$ day$^{-1}$}&{}\\
{\footnotesize $d_1$}&{\footnotesize death rate of hunting cells}&
{\footnotesize 0.0412 day$^{-1}$}&{\footnotesize \cite{Kusnetsov1994}}\\
{\footnotesize $q_2$ }&{\footnotesize growth rate of resting cells}&
{\footnotesize 0.0245 day$^{-1}$}&{\footnotesize \cite{Sarkar2008}}\\
{\footnotesize $\tau$ }&{\footnotesize time delay in conversion from resting 
cells to hunting cells}&{\footnotesize 45.6 day }&{\footnotesize \cite{Sarkar2008}}\\
{\footnotesize $K_2$}&{\footnotesize carrying capacity of resting cells}&
{\footnotesize 1 x ${10}^7$ cells}&{\footnotesize \cite{Sarkar2008}}\\
{\footnotesize ${\beta}_1$}&{\footnotesize conversion rate from}&{\footnotesize 6.2 x ${10}^{-9}$ }&{\footnotesize 
\cite{Kusnetsov1994}}\\  
{}&{\footnotesize  resting to hunting cells}&{\footnotesize cells$^{-1}$ day$^{-1}$}&{}\\
\end{tabular}
\label{tab1}
\end{table}

\begin{table}[hbt]
\centering
\caption{Parameters according to the literature.}
\begin{tabular}{ c | l | c }
\textbf{\small Parameter}& \textbf{\small Definition}& \textbf{\small Ref.}\\ 
\hline {\footnotesize $p_i$}&{\footnotesize predation 
coefficients of chemotherapic}&{\footnotesize \cite{Pinho2002}}\\  
{\footnotesize }&{\footnotesize agent on cells (C, H, R)}&{\footnotesize }\\  
{\footnotesize $a_i$}&{\footnotesize  determine the rate at which 
C, H, R,}&{\footnotesize \cite{Pinho2002}}\\
{\footnotesize }&{\footnotesize in the absence of competition and}&
{\footnotesize }\\
{}&{\footnotesize predation, reach carrying capacities}&{\footnotesize}\\
{\footnotesize $g_i$}&{\footnotesize represent the combination 
rates of the}&{\footnotesize \cite{Pinho2002}}\\  
{}&{\footnotesize chemotherapic agent with the cells}&{\footnotesize}\\
{\footnotesize $\Delta$}&{\footnotesize represents the infusion 
rate}&{\footnotesize \cite{Pinho2002}}\\  
{\footnotesize}&{\footnotesize of chemotherapy}&{\footnotesize }\\  
{\footnotesize $\xi$}&{\footnotesize washout rate of chemotherapy}&
{\footnotesize \cite{Pinho2002}}\\  
\end{tabular}
\label{tab2}
\end{table}

Introducing the following dimensionless variables
\begin{eqnarray} 
\bar{t}&=&\frac{t}{\rm day},\; \; \; \;  \bar{C} = \frac{C}{K_T},\; \; \; 
\; \bar{H} = \frac{H}{K_T}, \nonumber \\
\bar{R} &=& \frac{R}{K_T},\; \; \; \;
\bar{Z}=\frac{Z}{{\Delta}_M \,{\xi}^{-1}},\label{adimen}
\end{eqnarray}
where $K_T = K_1 + K_2$\ is the total carrying capacity and ${\Delta}_M$ is 
equal 1 mg ${\rm m}^{-2}{\rm day}^{-1}$. Combining (\ref{adimen}) with 
(\ref{modelo}), and relabelling the variables $\{\bar{t}$, $\bar{C}$, 
$\bar{H}$, $\bar{R}$, $\bar{Z}\}$ as t, C, H, R, Z, respectively, and the 
parameters $\{{\bar{q}}_1$, ${\bar{K}}_1$, ${\bar{\alpha}}_1$, ${\bar{p}}_1$, 
${\bar{g}}_1$, ${\bar{a}}_1$, ${\bar{\beta}}_1$, ${\bar{d}}_1$, 
${\bar{\alpha}}_2$, ${\bar{p}}_2$, ${\bar{g}}_2$, ${\bar{a}}_2$, ${\bar{q}}_2$, 
${\bar{K}}_2$, ${\bar{p}}_3$, ${\bar{g}}_3$, ${\bar{a}}_3$, $\bar{\Delta}$, 
$\bar{\xi}\}$ as $\{q_1$, ${K}_1$, ${\alpha}_1$, ${p}_1$, ${g}_1$, ${a}_1$, 
${\beta}_1$, $d_1$, ${\alpha}_2$, ${p}_2$, ${g}_2$, ${a}_2$, $q_2$,  ${K}_2$, 
${p}_3$, ${g}_3$, ${a}_3$, $\Delta$, $\xi\}$, respectively, we obtain the same 
equations for $C$, $H$ and $R$. However, the equation for $Z$ presents a small 
alteration,
\begin{eqnarray} 
\frac{dZ(t)}{dt} &=& \Delta\xi - \left( \xi +\frac{g_1 C(t)}{a_1 + C(t)} +
\frac{g_2 H(t)}{a_2 + H(t)} \right. \nonumber \\ 
&& + \left. \frac{g_3 R(t)}{a_3 + R(t)}\right) Z(t), 
\label{modeloadimensionalizado}
\end{eqnarray}
where we consider
\begin{eqnarray} 
{\bar{q}}_1 &=& q_1 \,{\rm day} ,\; \; \; \;  \bar{{\alpha}}_1 = {\alpha}_1 K_T 
\,{\rm day},\; \; \; \; {\bar{K}}_1 = \frac{K_1}{K_T} ,\; \; \; \; \nonumber \\
{\bar{p}}_1 &=& \frac{p_1 \,{\Delta}_M  \,{\rm day}}{K_T\, \xi},\;\;\; 
{\bar{a}}_1 =\frac{a_1}{K_T},\; \; \; \; {\bar{\beta}}_1 =  {\beta}_1 K_T \,
{\rm day},\nonumber \\  
{\bar{d}}_1  &=& d_1  \,{\rm day} ,\; \; \; \;  
\bar{{\alpha}}_2 = {\alpha}_2 K_T \,{\rm day} ,\;\;\; 
{\bar{g}}_1 = g_1 \,{\rm day} ,\nonumber \\
{\bar{g}}_2 &=& g_2 \,{\rm day} ,\; \; \; \; {\bar{g}}_3 = g_3 {\rm day} ,
\; \; \; \; {\bar{p}}_2 = \frac{p_2 \,{\Delta}_M  \,{\rm day}}{K_T\, \xi},
\nonumber  \\ 
{\bar{a}}_2 &=& \frac{a_2}{K_T}, \; \; \; \;{\bar{K}}_2 = \frac{K_2}{K_T} ,\; \;
\; \; {\bar{p}}_3 = \frac{p_3 \,{\Delta}_M {\rm day}}{K_T\, \xi}, \\  
{\bar{a}}_3 &=& \frac{a_3}{K_T}, \;\; \bar{\Delta} =\frac{\Delta}{\Delta_M}, 
\; \; {\bar{q}}_2 = q_2 \, {\rm day} , \; \; \bar{\xi} = \xi \, 
{\rm day}.\nonumber 
\label{parametrosadimen}
\end{eqnarray}

\begin{table}[hbt]
\centering
\caption{Dimensionless parameters.}
\begin{tabular}{ c | c | c | c }
\textbf{\small Parameter}& \textbf{\small Value}& \textbf{\small Parameter}& 
\textbf{\small Value}\\ \hline
{\footnotesize $q_1$}&{\footnotesize 0.18 }& {\footnotesize $K_1$}&
{\footnotesize 1/3 }\\
{\footnotesize ${\alpha}_1$}&{\footnotesize 1.6515 }&{\footnotesize 
${\alpha}_2$}&{\footnotesize 5.133 x ${10}^{-3}$}\\ 
{\footnotesize $d_1$}&{\footnotesize 0.0412}&{\footnotesize $q_2$ }&
{\footnotesize 0.0245 }\\
{\footnotesize $\tau$ }&{\footnotesize 45.6 }&{\footnotesize $K_2$}&
{\footnotesize 2/3}\\
{\footnotesize ${\beta}_1$}&{\footnotesize 9.3 x ${10}^{-2}$ }&{\footnotesize 
$p_1$}&{\footnotesize 1 x ${10}^{-3}$ }\\  
{\footnotesize $p_2$}&{\footnotesize 1 x ${10}^{-3}$  }&{\footnotesize $p_3$}&
{\footnotesize 1 x ${10}^{-3}$ }\\  
{\footnotesize $a_1$}&{\footnotesize 1 x ${10}^{-4}$ }&{\footnotesize $a_2$}&
{\footnotesize 1 x ${10}^{-4}$ }\\  
{\footnotesize $a_3$}&{\footnotesize 1 x ${10}^{-4}$ }&{\footnotesize $g_1$}&
{\footnotesize 0.1  }\\  
{\footnotesize $g_2$}&{\footnotesize 0.1}&{\footnotesize $g_3$}&{\footnotesize 
0.1 }\\  
{\footnotesize $\Delta$}&{\footnotesize 0 - ${10}^{4}$  }&{\footnotesize $\xi$}
&{\footnotesize  0.2 }\\  
\end{tabular}
\label{tab3}
\end{table}

\section{Continuous chemotherapy}

In this section we consider the continuous application of chemotherapy, 
without pause or interruption. That is, the value of the $\Delta$ is constant 
in time.

\subsection{Cancer suppression}
 
We consider the following initial conditions: ${C}_0  = 0.18$, ${H}_0 = 0.01$, 
${R}_0 = 0.48$ and $Z_0 = 0$. 
These initial conditions are in the limit cycle region of the model solution.
The periodic behaviour implies that the tumor levels oscillate around a fixed 
point, a clinically observed behaviour known as Jeff's phenomenon.
One cancer cell in the model (\ref{modelo}) is equal to $66 \times 10^{-9}$ in 
the dimensionless model. Our main aim is to find parameters that make the 
chemotherapy successfully suppress cancer, but that preserves the 
lymphocytes population. 

\begin{figure}
\begin{center}
\includegraphics[height=8cm,width=8cm]{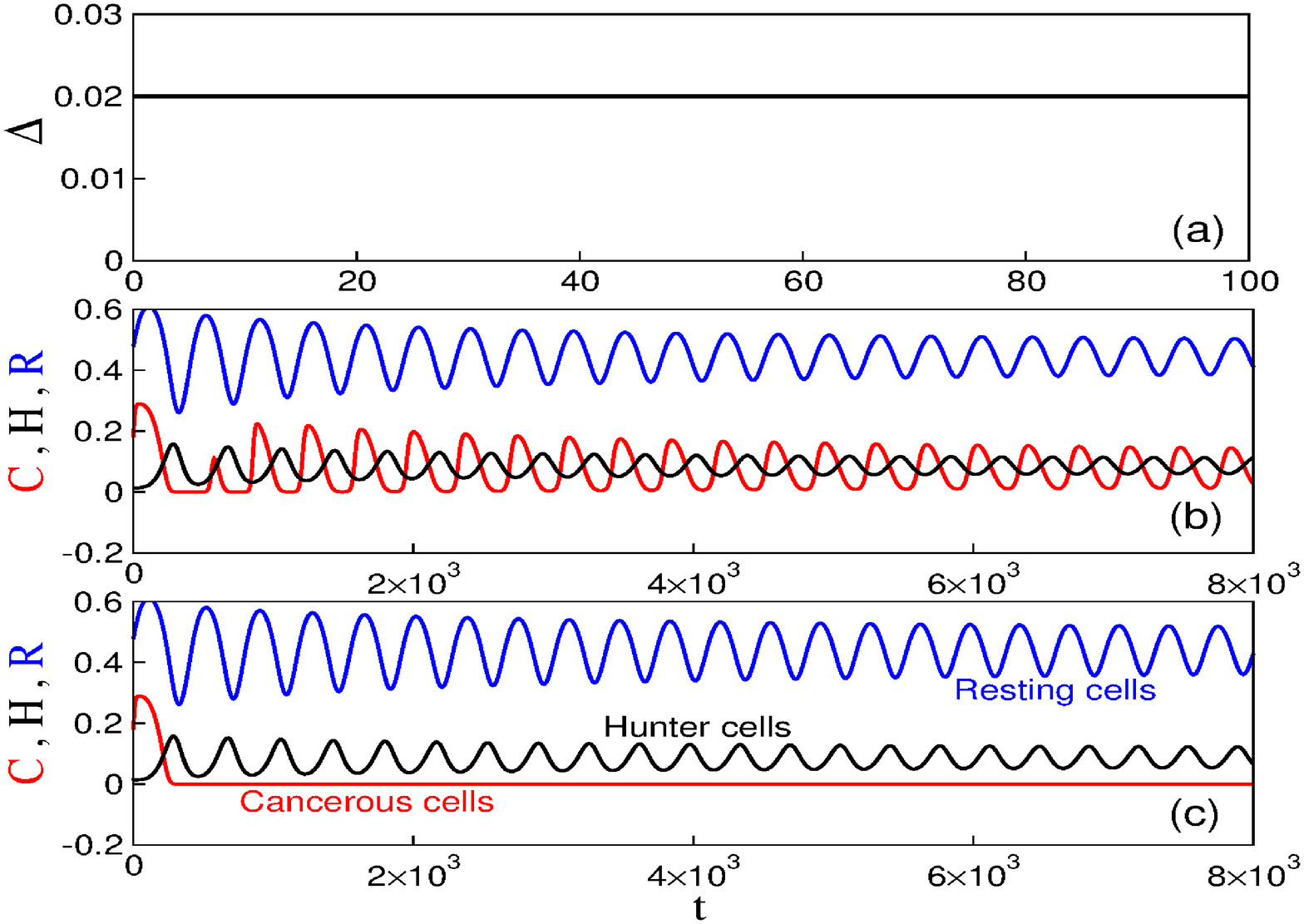}
\caption{(Color online) Time evolution of the dimensionless quantities 
according to the model (\ref{adimen}) using (\ref{modeloadimensionalizado}). 
(a) Continuous infusion rate of chemotherapy, (b) $\Delta=0.02$ and (c) cancer 
suppression considering $\Delta=0.025$. The red line represents the cancerous 
cells, black line the hunter cells and blue line the resting cells.}
\label{f1}
\end{center}
\end{figure}

\begin{figure}
\begin{center}
\includegraphics[height=5cm,width=7cm]{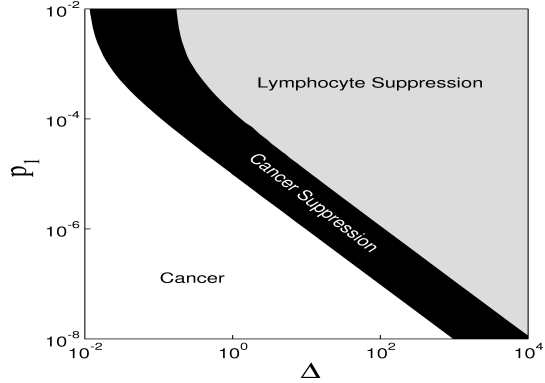}
\caption{Parameter space $p_1$ versus $\Delta$: the white region represents
parameters that lead to the existence of cancer cells, the black region 
to cancer suppression, and the grey region to the disappearance of the 
cancer cells and lymphocytes. We consider the continuous application of
chemotherapy, without pause or interruption.
}
\label{f2}
\end{center}
\end{figure}

Figure \ref{f1} displays the time evolution of the dimensionless quantities
and variables. Figure \ref{f1}(a) shows the behaviour of the infusion rate of 
chemotherapy. For $\Delta=0.02$ [Fig. \ref{f1}(b)] there is no cancerous 
suppression and the system presents stable oscillatory behaviour. If we 
increase the value of $\Delta$ to 0.025 [Fig. \ref{f1}(c)], the system may 
present cancerous suppression without the disappearance of lymphocytes. 
However, for larger $\Delta$, not only the cancer cells, but also 
lymphocytes disappear. To obtain a global picture of the parameters leading to
different behaviour of our model, we construct the parameter space shown in 
Figure \ref{f2}, for the parameters $\Delta$ and the predation coefficient of 
chemotherapy, $p_1$. For $p_1=0$ the rate of cancer cells proliferation is 
unaffected by the chemotherapic agent. This case may be interpreted as the use 
of  inappropriated drugs or mechanisms related to drug resistance. When $p_1$ 
is not null, there is drug-induced killing of cancer cells.
These parameters are important due to the fact that they
are directly related to the chemotherapy used in the treatment of cancer.
We can identify three behaviours in this parameter space. In the white region 
there are cancer cells. The suppression of cancer occurs for the parameters 
in the black region. The grey region presents an undesired situation, the 
suppression of lymphocytes. Therefore, we observe that it is possible to 
achieve cancerous suppression by increasing infusion rate of the chemotherapy 
to a high enough value, but the threshold depends on the value of $p_1$. For 
the parameters $\Delta$ and $p_1$ in the black region the number of malignant 
tumour cells goes to zero preserving the immune cells. When the number of 
cancer cells is zero, the treatment can stop and the tumour will not return.

\subsection{Lyapunov exponents}

\begin{figure}
\begin{center}
\includegraphics[height=9cm,width=9cm]{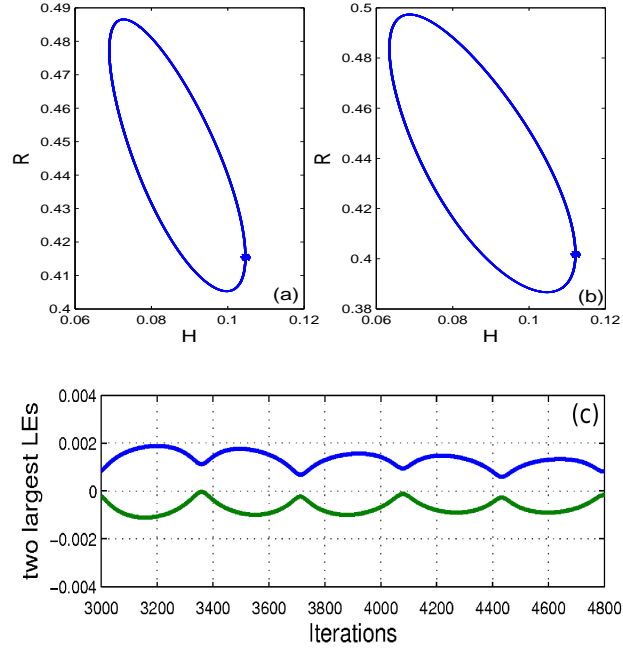}
\caption{Phase space plot for the variables $R$ and $H$, where the filled 
points is a Poincare map of the maximal values for the variable $H$. We 
consider the parameters according to Table \ref{tab3}, (a) $\Delta=0.01$ 
and (b) $\Delta=0.025$. (c) Time evolution for the two largest Lyapunov 
considering $\Delta=0.025$.}
\label{f3}
\end{center}
\end{figure}

To verify whether the suppression of cancer ($C=0$) is stable, that is, the 
cancer will not return after its elimination, we calculate the spectra of 
Lyapunov exponents of our model with time delay. We present only the values of 
the two largest Lyapunov exponents $\lambda_i(t)$ ($i=1,2$) at time $t$. We 
are interested in the maximal value of $\lambda_i(t)$. Firstly, we use  a 
value of $\Delta=0.01$, in that the therapy does not eliminate the cancer. 
In this case, the system oscillates in a stable limit cycle (Fig. \ref{f3}a). 
The largest Lyapunov exponent is about $0$ and the second largest negative. To 
determine that no cancer is an unstable solution of the model, we calculate 
the conditional Lyapunov exponents of the whole system but requiring the 
trajectory to lie in the subspace $C=0$. We obtain one positive Lyapunov 
exponent \cite{Baptista11}, which must be associated with the stability in 
this subspace since the Lyapunov exponents of the 3D reduced version of our 
model in (\ref{modeloadimensionalizado}), without the variable for $C$, are all 
negative. In other words, if $\Delta=0.01$, cancer will certainly not be 
eliminated.

For $\Delta=0.025$, not only the cancer can be eliminated, but also the 
dynamics in the subspace $C=0$ is stable. Figure \ref{f3}(b) shows that the 
behaviour of the system is a limit cycle. The result of the two largest 
Lyapunov exponents for the case $\Delta=0.025$ is given in Figure \ref{f3}(c). 
The maximum values $\lambda_1(t)$ and $\lambda_2(t)$, for $t>4400$, are 0.0015 
and -0.0001. Notice that this is only an upper bound for the real value and it 
indicates that the real value, whatever it is, needs necessarily to be smaller 
than 0.0015, a small number that we interpret as being 0, as required for a 
limit cycle.

\section{Pulsed chemotherapy}

Often chemotherapy treatments are carried out in cycles. The repeated 
application of drugs for a short time is a typical protocol for chemotherapy, 
called pulsed chemotherapy \cite{Pillis03}. For example, in this protocol, one 
may use the drug doxorubicin combined with other drugs to treat some types of 
cancer. The chemotherapy with these drugs is given through cycles of treatment 
according to the type of cancer \cite{Shulman}. 

In the following, we will consider two clinical protocols for the chemotherapy 
with respect to their dependence on the rate $\Delta$. One protocol is to 
administer the drug at a constant $\Delta$ and the other for two values of this
rate ($\Delta_1$ and $\Delta_2$), both of which have been applied with a 
determined period $P$ between two chemotherapic sessions. Our aim is to obtain 
the cancer suppression, while preserving the lymphocytes.

\subsection{First protocol}

\begin{figure}
\begin{center}
\includegraphics[height=8cm,width=8cm]{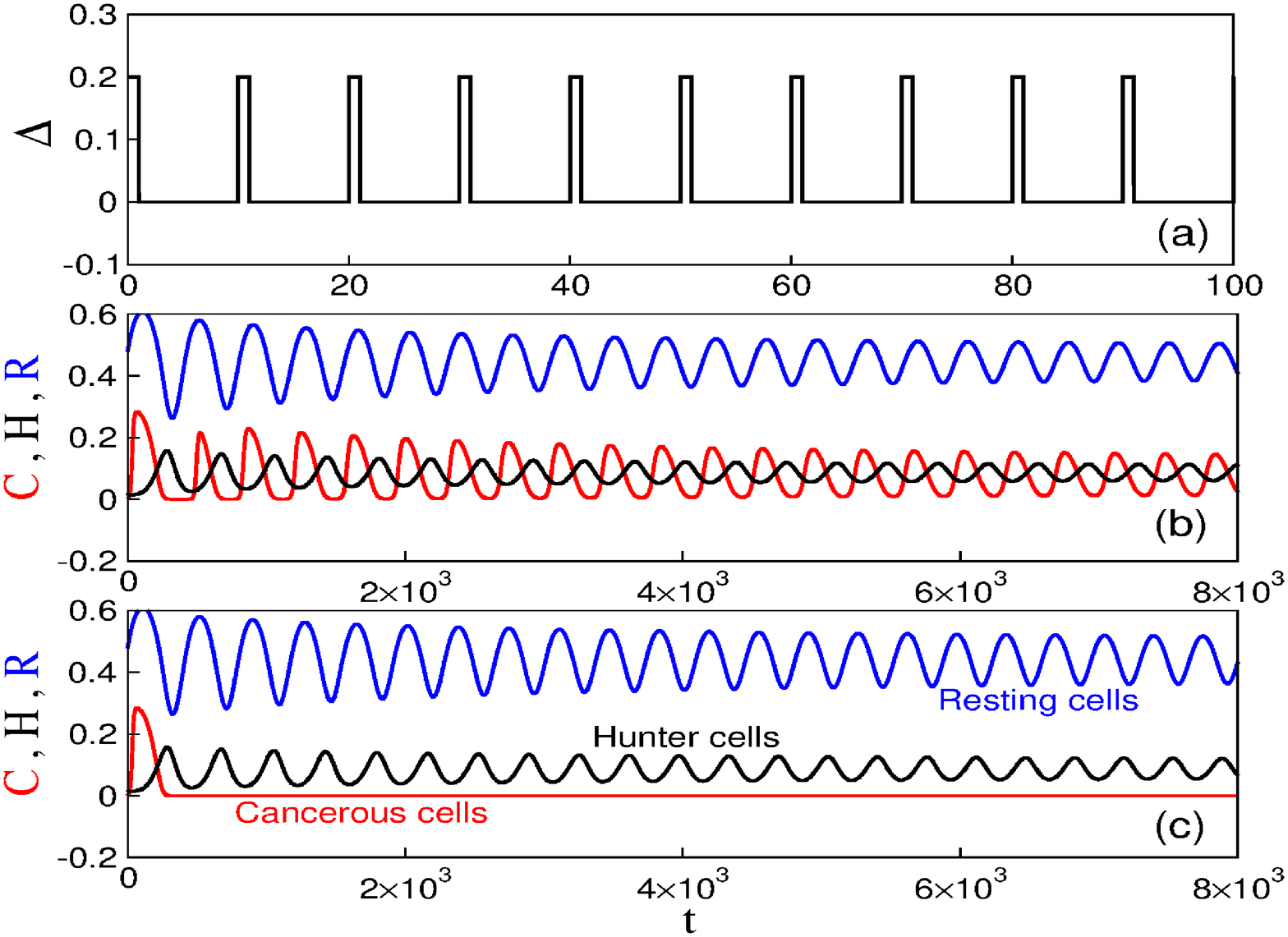}
\caption{(Color online) (a)$\Delta$ versus $t$, where $\Delta$ is not zero when 
the drug is applied with a period equal to 10 ($P=10$) with (b) $\Delta = 0.2$ 
and (c) $\Delta = 0.3$.}
\label{f4}
\end{center}
\end{figure}

Figure \ref{f4}(a) exhibits the drug injection pattern for the first protocol.
When $\Delta=0.2$, $P=10$, and parameters in Table \ref{tab3}, the tumour cell 
population does not vanish, as depicted in Figure \ref{f4}(b). However, 
considering the same period $P$, it is possible to obtain cancer suppression 
for $\Delta=0.3$ (Fig. \ref{f4}c).

\begin{figure}
\begin{center}
\includegraphics[height=5cm,width=7cm]{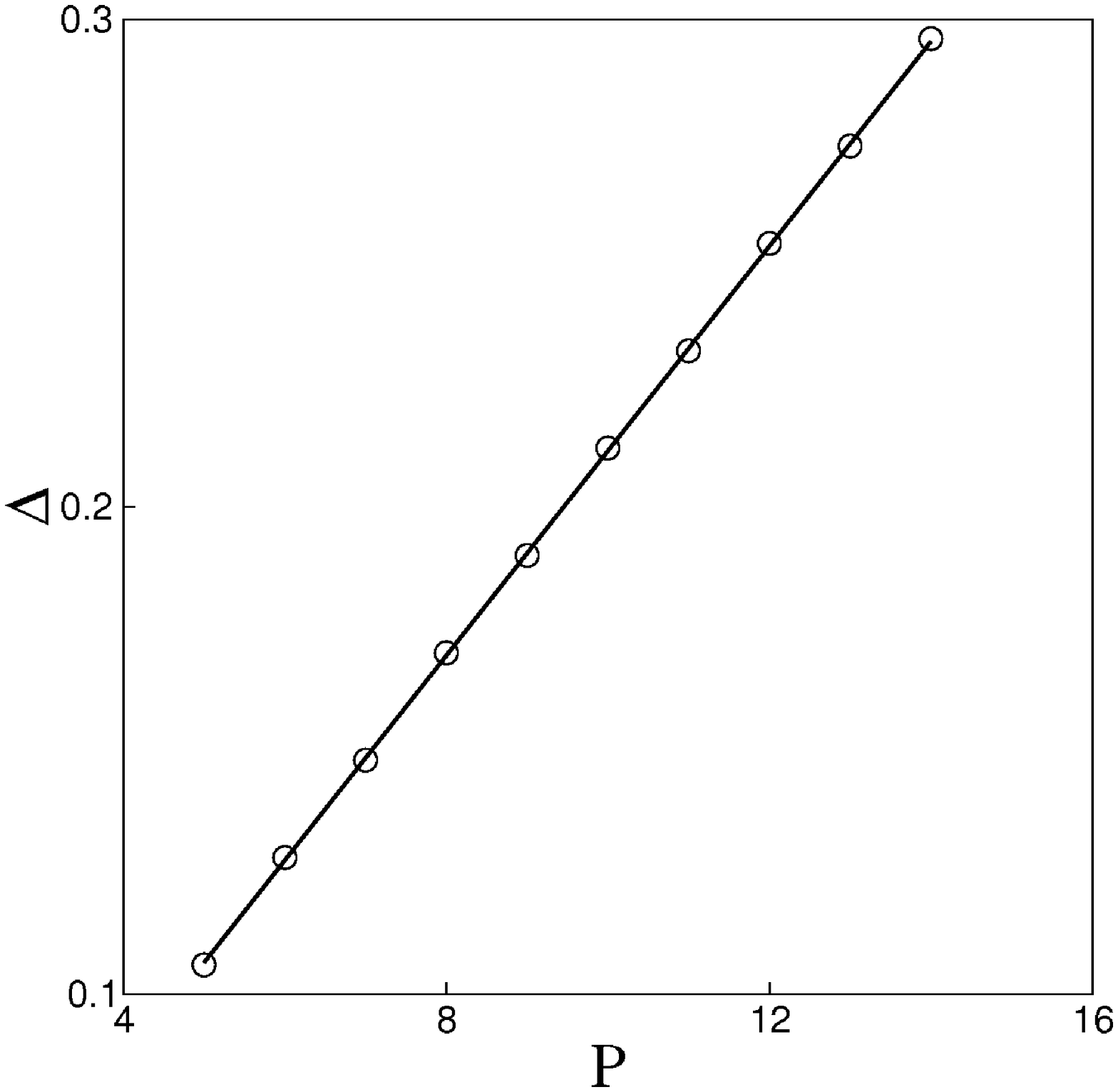}
\caption{$\Delta$ versus $P$, where the points represent the minimum value
of $\Delta$ for a given $P$ to occur the cancer suppression.}
\label{f5}
\end{center}
\end{figure}

The drugs have specific protocols of application according to the type of 
tumour. For this reason we study the period of the drug injection. Figure 
\ref{f5} exhibits the time interval $P$ of the pulsed chemo\-therapy, where the
points are used to denote the minimum value of the rate $\Delta$ in which the 
cancer suppression occurs. When $P$ increases, it is necessary to increase the 
intensity of the chemotherapy to obtain cancer suppression. As a matter of 
fact, the infusion rate versus the period shows a linear increase with the 
critical value of $\Delta$ as $P$ grows, $\Delta_c(P)\sim P$.

\begin{figure}
\begin{center}
\includegraphics[height=5cm,width=7cm]{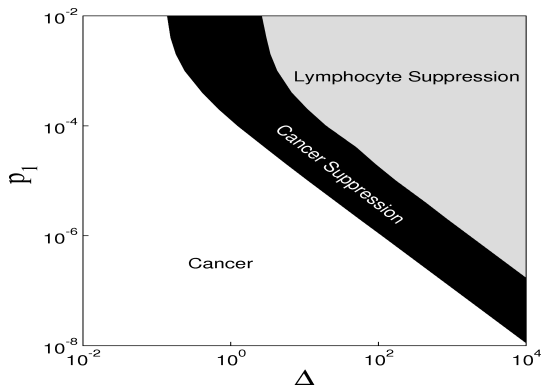}
\caption{Parameter space $p_1$ versus $\Delta$: the white region 
represents parameters that lead to the existence of cancer cells, the black 
region to cancer suppression, and the grey region to the disappearance of the 
cancer cells and lymphocytes. The drug is applied with a period equal to 10 
($P=10$).}
\label{f6}
\end{center}
\end{figure}

To verify the effect of the period on the behaviour of our model, we show the 
parameter space in Fig. \ref{f6}, similar to the one shown in Figure \ref{f2}, 
but considering pulsed chemotherapy with a period equal to 10. 
We also see the three behaviours: white regions represent parameters that
lead to cancer, black regions represent parameters that lead to the suppression
of cancer, and the grey region represent parameters that lead to the 
suppression of lymphocytes. Therefore, it is still possible to obtain a 
successful chemotherapy. However, the threshold of values of $p_1$ and 
$\Delta$ leading to a successful chemotherapy are larger in the pulsed
chemotherapy with $P=10$, then these threshold values for a continuous 
chemotherapy. This is a realistic behaviour observed in treatments that 
prescribe chemotherapic drugs in a continuous or pulsed way.

\subsection{Second protocol}

Another case for the pulsed chemotherapy consists in infusions 
of drugs with different concentrations and periods. There is recent research 
about the successful rate of each type of pulsed chemotherapy \cite{Kuebler}. 
For instance, the treatment for colon cancer adding oxaliplatin to bolus, 
fluorouracil mixed with leucovorin has been used with different infusion rates
\cite{Kuebler}. Due to clinical treatment described in the literature, we 
consider $\Delta$ to oscillate between two values.

\begin{figure}
\begin{center}
\includegraphics[height=8cm,width=8cm]{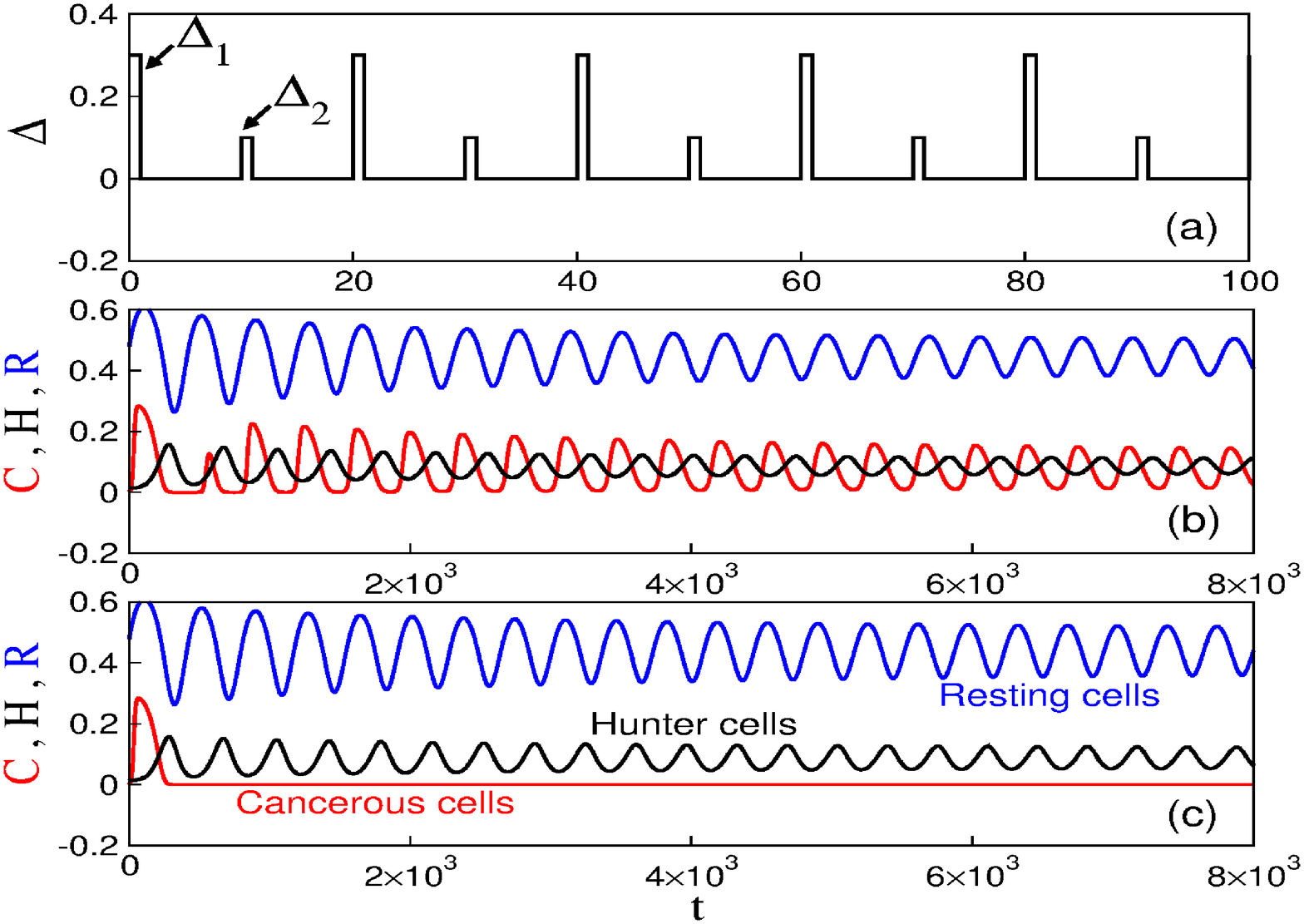}
\caption{(Color online) (a) $\Delta$ versus $t$, where  ${\Delta}_1= 0.3$ and 
${\Delta}_2= 0.1$. (b) Time evolution according to model 
(\ref{modeloadimensionalizado}) with period equal 10, ${\Delta}_1 = 0.3$ and 
${\Delta}_2 = 0.1$. (c) Applications of $P=10$ with ${\Delta}_1 = 0.3$ and 
${\Delta}_2 = 0.15$.}
\label{f7}
\end{center}
\end{figure}

Figure \ref{f7}(a) shows the drug injection pattern with $\Delta_1=0.3$,
$\Delta_2=0.1$, and $P=10$. For these values of $\Delta$ the cancer cells
do not disappear but they oscillate at regular intervals (Fig. \ref{f7}b). On 
the other hand, fixing $\Delta_1$ and increasing $\Delta_2$ the cancer is 
suppressed, as can be observed in Figure \ref{f7}(c).

\begin{figure}
\begin{center}
\includegraphics[height=5cm,width=7cm]{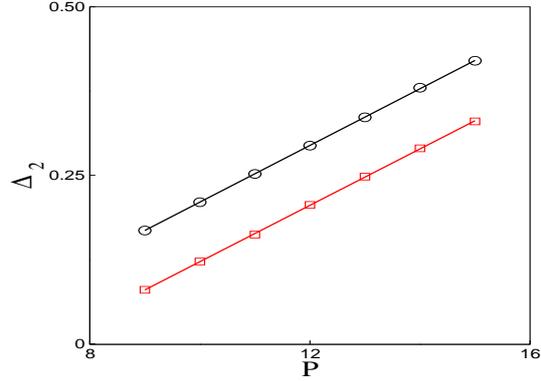}
\caption{$\Delta_2$ versus $P$, where the lines correspond to cancer 
suppression. $\Delta_1=0.212$ is for circles and $\Delta_1=0.3$ for squares.}
\label{f8}
\end{center}
\end{figure}

When the infusion rate is constant and chemotherapy sessions are periodically 
repeated, the cancer suppression depends on the time interval $P$. For this 
reason we fix $\Delta_1$ and vary $P$ and $\Delta_2$ that results in the
successful treatment of cancer. As a result, we see in Figure \ref{f8}, the 
lines represent the minimal values of $\Delta_2$ for a given $P$ that lead to 
cancer suppression. The circles are for $\Delta_1=0.212$ and the squares are 
for $\Delta_1=0.3$. Fixing $\Delta_2$ and varying $\Delta_1$, we obtain the 
same result.

Through the Figure 5 and Figure 8, we can see a linear relation between 
$\Delta$ and $P$. Defining the frequency as $f=1/P$, the infusion rate
versus the frequency will show a linear decreases with the critical value
of $\Delta$ as $f$ decreases, $\Delta_c(f)\sim f^{-1}$. This way, when
tuning the frequency, the value of the rate $\Delta$ in that the cancer
suppression occurs is altered.

The drug Diethylstilbestrol to prostate cancer can be used in treatment by
continuous or pulsed chemotherapy. In a continuous treatment may be administered
50 mg (oral) per day, every day, while in a pulsed treatment this drug may be 
administered 500 mg (infusion) once per week, many weeks. Comparing, we can 
see that the value of the pulsed is ten times the value of the continuous. 
Then, the values that we used are in accordance with realistic values.

\subsection{Exponential decline of cancer cells}

Michor and collaborators \cite{Michor} analysed 169 chronic mye\-loid leukaemia
patients, a cancer of the white blood cell, or leukocytes. They studied
the dynamics of different treatment responses to tyrosine kinase inhibitor
imatinib, which is also known as glivec. The imatinib is used to treat some 
types of leukaemia and soft tissue sarcoma. The treatment with this drug 
causes the death of cancer cells by inhibiting the signals exchanged by the 
cancer cells responsible to produce the growth and the division of the cancer
cells.

\begin{figure}
\begin{center}
\includegraphics[height=6cm,width=7cm]{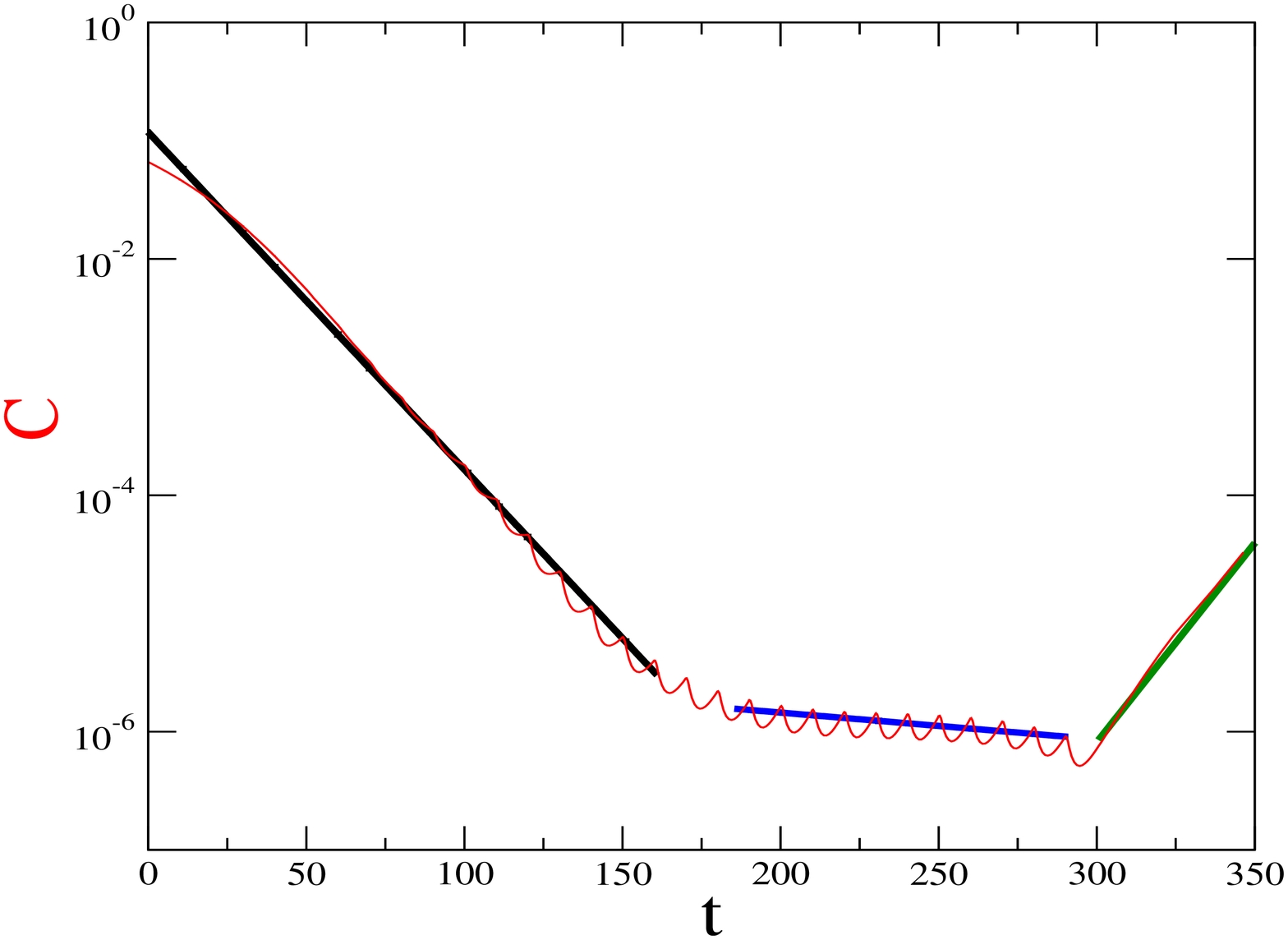}
\caption{(Color online) Time evolution of the dimensionless quantity $C$ 
according to the model (\ref{adimen}) using (\ref{modeloadimensionalizado}), 
where we consider $P=10$ and $\Delta=0.3$. The value of the first slope is 
$-0.06$ (black line), the second is $-0.005$ (blue line), and the third is 
$0.08$ (green line).}
\label{f9}
\end{center}
\end{figure}

In Ref. \cite{Michor} was showed that a successful therapy using imatinib 
leads to a biphasis exponential decline of leukaemic cells in time. The value 
of the first slope (quantifying the exponential time decay rate of cancer 
cells) is approximately $-0.05$, and represents the death of leukaemic 
differentiated cells. The second slope is around $-0.008$, and is due to the 
death of leukaemic progenitors. If the imatinib therapy is interruped the 
slope is approximately $0.09$. Without imatinib the differentiated leukaemic 
cells arise from leukaemic stem cells. Figure \ref{f9} shows the slopes 
obtained through the model (\ref{adimen}). We consider $P=10$, $\Delta=0.3$, 
and two values for $p_1$. We use for the time interval $0<t\leq 175$ that
produces the first slope (black line) $p_1=0.001$, and for the time interval 
$175<t\leq 300$ that produces the second slope (blue line) $p_1=0.0009$. 
For $t>300$, time interval that produces the third slope (green line) there is 
not chemotherapy. As a result, we obtain the slopes $-0.06$, $-0.005$, and 
$0.08$, which are remarkably similar with the slopes obtained in Ref. 
\cite{Michor}.

In our model, the case without therapy can be simulated by setting the 
parameter $p_i$ equal to zero. This parameter can be also used to simulate 
a case where the cancer cells become resistante to the drug. The leukaemic 
cells may present resistance to imatinib therapy. Resistance can be modelled 
decreasing the effect of chemotherapy to cancer cells (i.e., decreasing $p_1$),
while leaving the effect on immune cells constant (maintaining $p_2$ and $p_3$).

\section{Conclusions}

We propose a delay differential equations model for the evolution of cancer 
under the attack of both the immune system and chemotherapy. The novelty in 
this model is the introduction of the chemotherapy and the adjustment of 
parameters according to recent experimental evidence. We considered some types 
of protocols aiming at the cancer suppression.

We studied a continuous administration of drugs. The solutions of the system 
are stable, presenting a limit cycle behaviour. We identified domains of cancer 
suppression for a wide parameter range of the predation coefficient $p_1$ 
of the chemotherapic agent and of the continuous infusion rate of chemotherapy, 
$\Delta$. Our main results in this session was to show that (i) $p_1$ and 
$\Delta$ that lead to a successful cancer treatment (elimination of cancer 
cells) are inversely related; (ii) too large values of $p_1$ and $\Delta$ 
eliminate cancer but also eliminate the lymphocytes.

The success of the chemotherapic treatment is highly dependent on the values
of the parameters $p_1$ and $\Delta$, responsible for the interactions
between the tumour and the drug. By varying $p_1$ we were able to obtain the 
biphasic exponential decline observed in chronic myeloid leukaemia
\cite{Michor}. Moreover, the variation of $p_1$ permits to simulate cases in 
which the tumour develops drug resistance.

We examined the behaviour of the cancer cells with pulsed chemotherapy and
its dependence on the chemo\-the\-rapic dosing regime. We verified the 
possibility of cancer suppression through two clinical protocols. In fact,
we investigated infusions of drugs with equal and different concentrations.
The protocol with different concentrations is more used in the case of drug 
combination. Our results enabled us to predict the values of the relevant 
parameters for cancer suppression through protocols related to the treatment 
of ill people.

\section*{Acknowledgements}

This study was partially supported by the following Brazilian Government 
Agencies: CNPq, CAPES, FAPESP and Funda\c c\~ao Arauc\'aria. HPR, MSB, CG
acknowledge the RSE-NSFC (443570/NNS/INT-61111130122). K. C. Iarosz 
acknowledges CAPES Foundation, Ministry of Education of Brazil, Bras\'ilia - 
Processo n. 1965/12-3.

\end{document}